\documentclass[twocolumn,english,aps,pra,showpacs]{revtex4}
\usepackage[T1]{fontenc}
\usepackage[latin1]{inputenc}
\usepackage{amsmath}
\usepackage{graphicx}
\usepackage{amssymb}
\usepackage{esint}

\makeatletter
\@ifundefined{textcolor}{}
{%
 \definecolor{BLACK}{gray}{0}
 \definecolor{WHITE}{gray}{1}
 \definecolor{RED}{rgb}{1,0,0}
 \definecolor{GREEN}{rgb}{0,1,0}
 \definecolor{BLUE}{rgb}{0,0,1}
 \definecolor{CYAN}{cmyk}{1,0,0,0}
 \definecolor{MAGENTA}{cmyk}{0,1,0,0}
 \definecolor{YELLOW}{cmyk}{0,0,1,0}
 }

\@ifundefined{definecolor}{\@ifundefined{definecolor}{\@ifundefined{definecolor}{\@ifundefined{definecolor}
 {\usepackage{color}}{}
}{}}{}}{}\makeatother

\makeatother

\usepackage{babel}

\makeatother

\usepackage{babel}

\makeatother

\usepackage{babel}

\makeatother

\usepackage{babel}

\makeatother

\usepackage{babel}

\makeatother

\usepackage{babel}

\begin{document}

\title{Many-body theories of density response for a strongly correlated
Fermi gas}

\author{Hui Hu}

\affiliation{ARC Centre of Excellence for Quantum-Atom Optics,\\
 Centre for Atom Optics and Ultrafast Spectroscopy,\\
 Swinburne University of Technology, Melbourne 3122, Australia}

\date{\today}
\begin{abstract}
Recent breakthroughs in the creation of ultra-cold atomic gases in
the laboratory have ushered in major changes in physical science.
Many novel experiments are now possible, with an unprecedented control
of interaction, geometry and purity. Quantum many-body theory is facing
severe challenges in quantitatively understanding new experimental
results. Here, we review some recently developed theoretical techniques
that provide successful predictions for density response of a strongly
correlated atomic Fermi gas. These include the strong-coupling random-phase
approximation theory, high-temperature quantum virial expansion, and
asymptotically exact Tan relations applicable at large momentum. 
\end{abstract}

\pacs{05.30.Jp, 03.75.Mn, 67.85.Fg, 67.85.Jk}

\maketitle

\section{Introduction}

Highly correlated systems of fermions are of great interest in a wide
range of fields ranging from astrophysics, nuclear physics, condensed
matter physics to atomic physics \cite{BlochRMP,GiorginiRMP}. This
great interest comes with an important generic idea of fermionic universality
\cite{HoUniversality,HDLNaturePhysics}. Any dilute Fermi gases with
sufficiently strong interactions should behave identically on a scale
given by the average particle separation, independent of the details
of the short-ranged interactions. Qualitatively, in a two-component
Fermi gas the fermionic universality emerges when the $s$-wave scattering
length $a$, which characterizes elastic collisions, is much large
than the mean interparticle spacing $n^{-3}$ and the range of the
scattering potential $r_{0}$ \cite{HoUniversality}, i.e., $a\gg n^{-3}\gg r_{0}$.
Using the Fermi wave-vector $k_{F}=(3\pi^{3}n)^{-3}$, this is equivalent
to the dimensionless condition of $k_{F}a\gg1\gg k_{F}r_{0}$.

The recently realized ultracold Fermi gases of $^{6}$Li and $^{40}$K
atoms near a broad collisional (Feshbach) resonance appear to be the
most appealing systems to study fermionic universality \cite{ThomasScience2002}.
By taking advantage of the ability to accurately create strong atom-atom
attractions with an external magnetic field, the crossover from a
Bose-Einstein condensate (BEC) of molecules (atom pairs) to a Bardeen-Cooper-Schrieffer
(BCS) superfluid of atoms has now been routinely demonstrated \cite{BlochRMP,GiorginiRMP}.
At the cusp of the crossover, where the $s$-wave scattering length
diverges ($a\rightarrow\pm\infty$), atomic Fermi gases should exhibit
universal behavior \cite{HoUniversality}. The Fermi gas in this limit
is referred to as unitary Fermi gas, since its scattering amplitude
is unitarily limited. In the extreme BEC and BCS limits, the elementary
constitutes of the system are clearly bosons and fermions, respectively.
However, in a unitary Fermi gas, bosonic and fermionic excitations
play an equally important role. To date, many properties of this new
state of matter haven been investigated experimentally in harmonically
trapped configurations, such as the pairing energy \cite{ChinScience2003},
the hydrodynamic expansion \cite{ThomasScience2002}, the frequency
of collective excitations \cite{ThomasPRL2004,GrimmPRL2004,HuPRL2004},
the condensate fraction \cite{JinPRL2004,KetterlePRL2004}, the vortex
lattice \cite{KetterleNature2005}, and the thermodynamic equation
of state \cite{JILAEoS,dukeEoS,ensEoS,HLDEoS}.

These novel and important experimental results impose severe challenges
for theorists. For a strongly correlated Fermi gas, there is no small
interaction parameter to set the accuracy of many-body theories \cite{HLDEoS,HLDcmpPRA2008}.
In recent years, great efforts have been given to develop better quantum
Monte Carlo simulations \cite{astrakharchik,bulgac,burovski,akkineni,carlson}
and strong-coupling theories \cite{OhashiPRL2002,HLDEPL2006,haussmann}.
A number of static properties, particularly the thermodynamic behavior,
have now been understood quantitatively \cite{HLDEoS}. However, dynamic
properties such as the single-particle spectral function \cite{JinNature2008}
and the dynamic density response \cite{BraggSwin}, are not as well
understood.

In this paper, we review our recent theoretical progress on understanding
the dynamic density response of a strongly correlated atomic Fermi
gas. The dynamic density response is characterized by the so-called
dynamic structure factor (DSF), which gives the linear response of
the many-body system to an excitation process that couples to density
\cite{allanbook,FetterBook}. For ultracold atomic gases, it can be
conveniently measured by two-photon Bragg spectroscopy using two laser
beams \cite{BraggSwin,BraggBEC}. In atomic BECs, density response
at small momentum has been measured to characterize the fundamental
Bogoliugbov phonon excitations \cite{BraggBEC}. In ultracold Fermi
gases, Bragg spectroscopy has been used to obtain dynamic structure
factor over the BEC-BCS crossover \cite{BraggSwin}, albeit at much
large momentum.

Our theoretical developments include both perturbative and non-perturbative
techniques. At low temperatures, we attempt the conventional random-phase
approximation theory \cite{ourRPA}, which is extended to the strongly
interacting crossover regime. At high temperatures, we develope a
novel quantum virial expansion method \cite{ourVirialDSF}. At large
momentum, we derive asymptotically exact results for the structure
factor \cite{ourTanSSF}, based on the well-known Tan relations \cite{TanRelations}.
All these theories lead to useful predictions that have been quantitatively
confirmed by Bragg spectroscopy measurements \cite{TanSwin1,TanSwin2,TanSwin3}.

This review is structured as follows. In the next section, we define
DSF and discuss briefly the existing experimental Bragg spectroscopy
results. In Sec. III, we introduce the random-phase approximation
theory and present the comparison of theory to experiment at low temperatures.
In Sec. IV, we describe the quantum virial expansion method and compare
the theoretical predictions with experimental data for a unitary Fermi
gas at high temperature. In Sec. V, we derive exact Tan relations
for the structure factor and explain the experimental confirmation
of these relations. The last section (Sec. VI) is devoted to the conclusions
and some final remarks.

\section{Dynamic structure factor and Bragg spectroscopy}

The DSF $S({\bf q},\omega)$ is the Fourier transform of the density-density
correlation functions at two different space-time points \cite{allanbook,FetterBook}.
For a two-component atomic Fermi gas with equal spin populations $N/2$
(referred to as spin-up, $\sigma=\uparrow$, and spin-down, $\sigma=\downarrow$),
we have $S_{\uparrow\uparrow}({\bf q},\omega)=S_{\downarrow\downarrow}({\bf q},\omega)$
and $S_{\uparrow\downarrow}({\bf q},\omega)=S_{\downarrow\uparrow}({\bf q},\omega)$,
each of which is defined by, \begin{eqnarray}
S_{\sigma\sigma^{\prime}}({\bf q},\omega) & = & Q^{-1}\sum_{nn^{\prime}}e^{-\beta E_{n^{\prime}}}\left\langle n\left|\delta\rho_{\sigma}\left({\bf q}\right)\right|n^{\prime}\right\rangle \times\nonumber \\
 &  & \left\langle n^{\prime}\left|\delta\rho_{\sigma^{\prime}}^{\dagger}\left({\bf q}\right)\right|n\right\rangle \delta\left(\hbar\omega-E_{nn^{\prime}}\right),\end{eqnarray}
 where $\left|n\right\rangle $ and $E_{nn^{\prime}}=E_{n}-E_{n^{\prime}}$
are, respectively, the eigenstate and eigenvalue of the many-body
system, while $Q=\sum_{n}\exp(-\beta E_{n})\equiv\sum_{n}\exp(-E_{n}/k_{B}T)$
is the partition function. The density operator $\delta\hat{\rho}_{\sigma}({\bf q})=\sum_{i\sigma}e^{-i{\bf q\cdot r}_{i}}$
is the Fourier transform of the atomic density operator $\delta\hat{\rho}_{\sigma}\left({\bf r}\right)$
for spin-$\sigma$ atoms. The total DSF is given by $S({\bf q},\omega)\equiv2[S_{\uparrow\uparrow}({\bf q},\omega)+S_{\uparrow\downarrow}({\bf q},\omega)]$.
The DSF satisfies two remarkable $f$-sum rules, \begin{equation}
\int_{-\infty}^{+\infty}S({\bf q},\omega)\omega d\omega=N\frac{\hbar{\bf q}^{2}}{2m}\end{equation}
 and \begin{equation}
\int_{-\infty}^{+\infty}S_{\uparrow\downarrow}({\bf q},\omega)\omega d\omega=0,\end{equation}
 where $m$ is the mass of atoms. The exact $f$-sum rule is important
for understanding interacting many-body systems, as it holds irrespective
of statistics and temperature.

According to the finite-temperature quantum field theory \cite{allanbook},
it is convenient to calculate DSF from dynamic susceptibility, $\chi_{\sigma\sigma^{\prime}}\left({\bf q},\tau\right)\equiv-\left\langle T_{\tau}\hat{\rho}_{\sigma}\left({\bf q},\tau\right)\hat{\rho}_{\sigma^{\prime}}\left({\bf q},0\right)\right\rangle $,
where $\tau$ is an imaginary time in the interval $0<\tau\leq\beta=1/k_{B}T$.
The Fourier component $\chi_{\sigma\sigma^{\prime}}\left({\bf q},i\omega_{n}\right)$
at discrete Matsubara imaginary frequencies $i\omega_{n}=i2n\pi k_{B}T$
($n=0,\pm1,...$) gives directly the DSF, after taking analytic continuation
and using the fluctuation-dissipation theorem: \begin{equation}
S_{\sigma\sigma^{\prime}}\left({\bf q,}\omega\right)=-\frac{\mathop{\rm Im}\chi_{\sigma\sigma^{\prime}}\left({\bf q};i\omega_{n}\rightarrow\omega+i0^{+}\right)}{\pi(1-e^{-\beta\omega})}\,\,\,.\label{fluct-dissi-theorem}\end{equation}

The frequency integral of the DSF defines the so-called static structure
factor (SSF). For different spin components, we have, \begin{equation}
S_{\sigma\sigma^{\prime}}\left({\bf q}\right)=\frac{2}{N}\int_{-\infty}^{+\infty}S_{\sigma\sigma^{\prime}}\left({\bf q,}\omega\right)d\omega.\end{equation}
 The total SSF is given by, $S\left({\bf q}\right)=(1/N)\int_{-\infty}^{+\infty}d\omega S\left({\bf q,}\omega\right)=S_{\uparrow\uparrow}({\bf q})+S_{\uparrow\downarrow}({\bf q})$.
The SSF is related to the two-body pair correlation function $g_{\sigma\sigma^{\prime}}\left({\bf r}\right)$
\cite{FetterBook}. For the spin-antiparallel SSF, the relation is,
\begin{equation}
S_{\uparrow\downarrow}\left({\bf q}\right)=\frac{N}{2}\int d{\bf r}\left[g_{\uparrow\downarrow}\left({\bf r}\right)-1\right]e^{i{\bf q\cdot r}}\text{.}\end{equation}

Experimentally, the DSF is measured by inelastic scattering experiments
of two-photon Bragg spectroscopy \cite{BraggSwin,BraggBEC}. The atoms
are exposed to two laser beams with differences in wave-vector and
frequency. In a two-photon scattering event, atoms absorb a photon
from one of the beams and emit a photo into the other. Therefore,
the difference in the wave-vectors of the beams defines the momentum
transfer $\hbar{\bf q}$, while the frequency difference defines the
energy transfer $\hbar\omega$. In the regime of large transferred
momentum, which is exactly the case in current experiments for the
crossover Fermi gas \cite{BraggSwin}, the single-particle response
is dominant and peaks at the quasi-elastic resonance frequency $\omega_{res}=\hbar{\bf q}^{2}/(2M)$,
where $M$ is the mass of the elementary constituents of the system.
Therefore, we may anticipate that the Bragg response peaks at $\omega_{R}=\hbar{\bf q}^{2}/(2m)$
in the BCS limit and peaks at $\omega_{R,mol}=\hbar{\bf q}^{2}/(4m)=\omega_{R}/2$
in the BEC limit, since the underlying particles are respectively
free atoms ($M=m$)\ and molecules ($M=2m$).

\begin{figure}[htp]

\begin{centering}
\includegraphics[clip,width=0.45\textwidth]{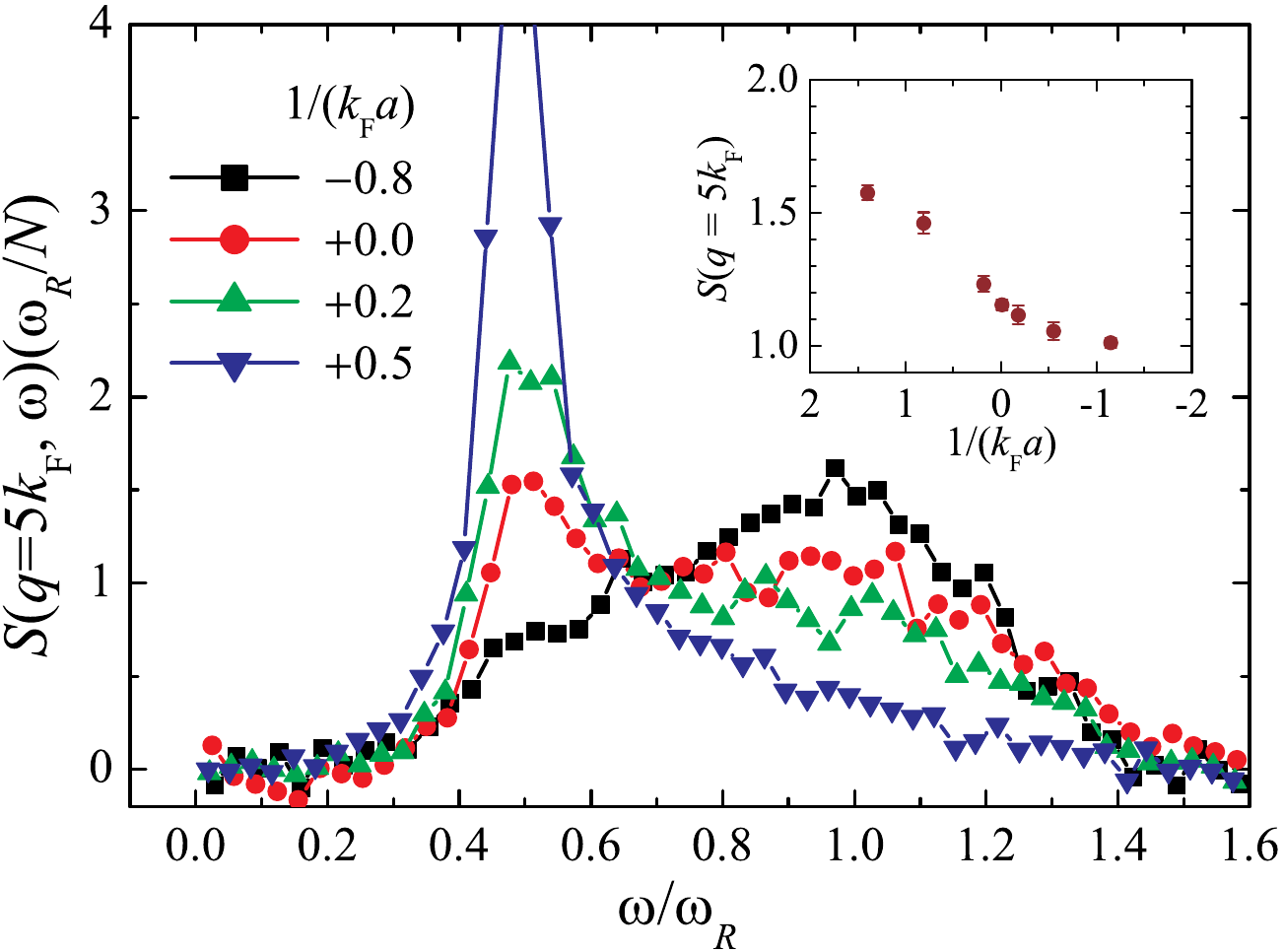} 
\par\end{centering}

\caption{(color online) Measured dynamic structure factor of a harmonically
trapped Fermi gas in the BEC-BCS crossover at the lowest attainable
temperature ($T<0.1T_{F}$) and at a large transferred wave-vector
$q=5k_{F}$. The inset shows the static structure factor as a function
of the dimensionless interaction parameter. The figure is reproduced
from ref. \cite{BraggSwin} with permission.}

\label{fig1} 
\end{figure}

\begin{figure}[htp]

\begin{centering}
\includegraphics[clip,width=0.45\textwidth]{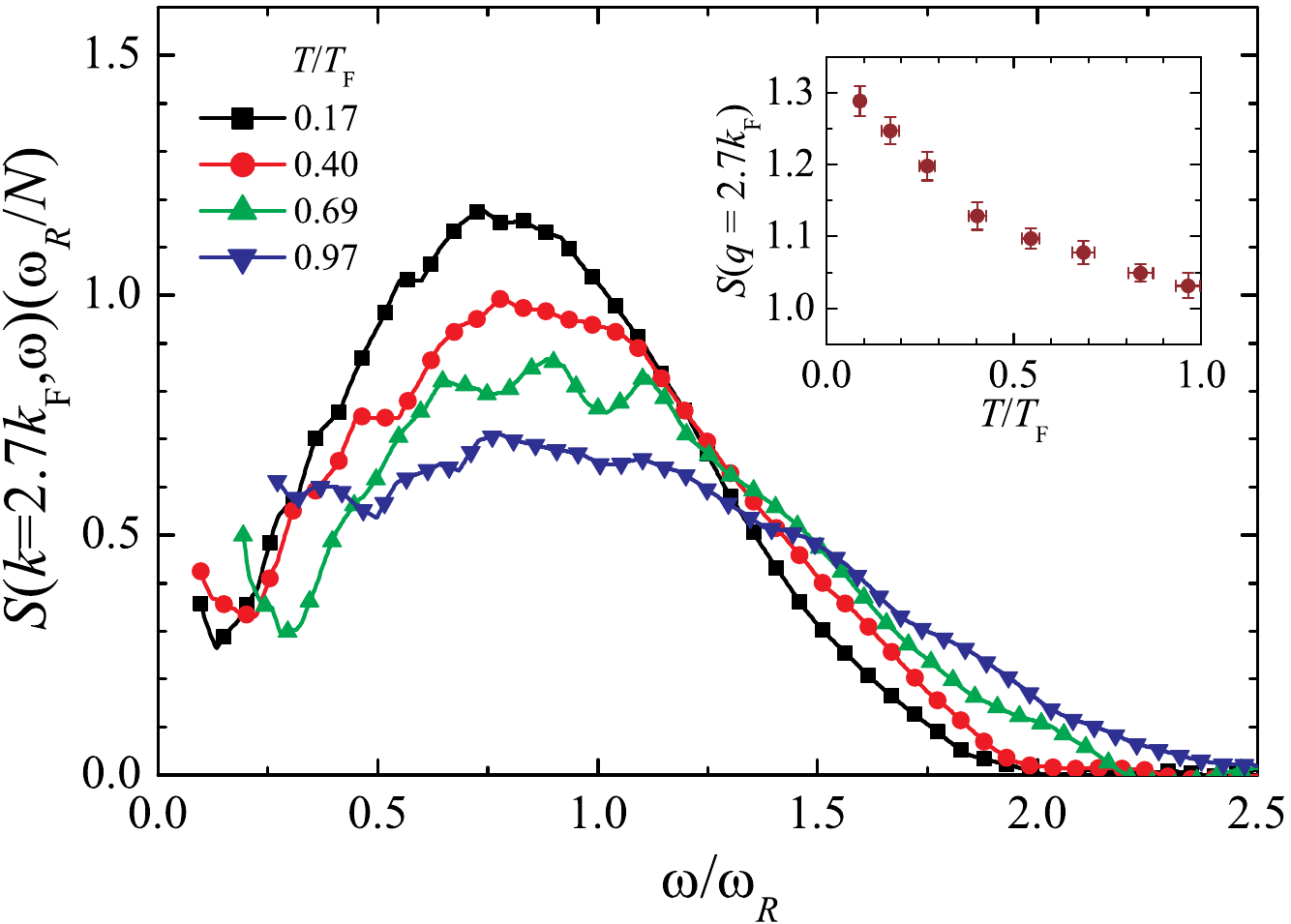} 
\par\end{centering}

\caption{(color online) Temperature dependence of dynamic structure factor
of a trapped Fermi gas in the unitary limit, measured at $q=2.7k_{F}$.
The inset shows the temperature dependence of static structure factor.
The figure is reproduced from refs. \cite{TanSwin2} and \cite{TanSwin3}
with permission.}

\label{fig2} 
\end{figure}

In Figs. 1 and 2, we summarize the main experimental results for a
harmonically trapped Fermi gas in the BEC-BCS crossover \cite{BraggSwin,TanSwin1,TanSwin2,TanSwin3}.
Fig. 1 shows the DSF (main panel) and SSF (inset) at several dimensionless
interaction strengths and at the lowest experimentally attainable
temperature (i.e., $T<0.1T_{F}$, where $T_{F}$ is the Fermi temperature)
\cite{BraggSwin}, while Fig. 2 presents the temperature dependence
of structure factors in the most interesting unitary limit \cite{TanSwin2,TanSwin3}.
As anticipated, in Fig. 1 the DSF peaks at $\omega_{R}/2$ and $\omega_{R}$
on the BEC side (i.e., $1/(k_{F}a)=+0.5$) and on the BCS side ($1/(k_{F}a)=-0.8$),
respectively. In the unitary limit, where the statistics of the elementary
excitations is not well defined, we observe a two-peak structure with
responses from both molecules and free-atoms. As the temperature increases
(Fig. 2), however, these two peaks emerge. The resultant broad peak
shifts eventually to $\omega_{R}$ at high temperatures.

The appearance of a molecular response in the BEC-BCS crossover is
also captured by the SSF. As shown in the insets of both figures,
with increasing interaction strength or temperature, the SSF increases
from $1$ to $2$. This can be understood from the $f$-sum rule.
By assuming that the response is exhausted by a single-mode excitation
at $\omega_{res}=\hbar{\bf q}^{2}/(2M)$, the $f$-sum rule can be
rewritten by, \begin{equation}
N\frac{\hbar{\bf q}^{2}}{2m}=\int_{-\infty}^{+\infty}S({\bf q},\omega)\omega d\omega\approx NS\left({\bf q}\right)\omega_{res}.\end{equation}
 Therefore, we obtain $S\left({\bf q}\right)=M/m$, which should change
from 1 to 2 through the crossover. In other words, we may regard $mS\left({\bf q}\right)$
as the effective mass of the elementary constituents of the system.

\section{Strong-coupling random-phase approximation theory}

In this section, we extend a perturbative random-phase approximation
(RPA) theory to strongly interacting regime and use it to describe
{\em quantitatively} the observed low-temperature Bragg spectra
for harmonically {\em trapped} $^{6}$Li atoms at large transferred
momenta (see Fig. 1) \cite{ourRPA}. The RPA method has previously
been applied to study the DSF \cite{minguzzi} and collective oscillations
\cite{bruun} of weakly interacting Fermi superfluids. A dynamic mean-field
approach \cite{combescot,combescotPRA}, identical to the RPA but
based on kinetic equations, was developed to investigate structure
factors and collective modes of a uniform, strongly interacting Fermi
gas.

\subsection{The basic idea of RPA theory}

To outline briefly the central idea of RPA, we consider the single-channel
Hamiltonian \cite{LHPRA2005},

\begin{eqnarray}
{\cal H} & = & \sum_{\sigma}\int d{\bf r}\psi_{\sigma}^{+}({\bf r})\left[-\frac{\hbar^{2}\nabla^{2}}{2m}-\mu+V_{T}({\bf r})\right]\psi_{\sigma}({\bf r})\nonumber \\
 &  & +U_{0}\int d{\bf r}\psi_{\uparrow}^{+}({\bf r})\psi_{\downarrow}^{+}({\bf r})\psi_{\downarrow}({\bf r})\psi_{\uparrow}({\bf r}),\label{hami}\end{eqnarray}
 which describes a balanced spin-1/2 Fermi gas in a harmonic trap
$V_{T}({\bf r})$, where fermions with unlike spins interact via a
contact potential $U_{0}\delta({\bf r-r}^{\prime})$. The total number
of atoms $N$ is tuned by the chemical potential $\mu$ and the bare
interaction strength $U_{0}$ is renormalized by the \textit{s}-wave
scattering length, \begin{equation}
\frac{1}{U_{0}}+\sum_{{\bf k}}\frac{m}{\hbar^{2}{\bf k}^{2}}=\frac{m}{4\pi\hbar^{2}a}.\end{equation}
 In the superfluid phase, we treat the system as a gas of long-lived
Bogoliubov quasiparticles interacting through a mean-field and consider
its response to a weak external field of the form of $\delta Ve^{i({\bf q\cdot r}-\omega t)}$.
The essential idea of the RPA is that there is a self-generated mean-field
potential experienced by quasiparticles, associated with the local
changes in the density distribution of the two spin species, $\delta U=U_{0}\int d{\bf r(}\sum_{\sigma}\delta n_{\sigma}\psi_{\sigma}^{+}\psi_{\sigma}+\delta m\psi_{\uparrow}^{+}\psi_{\downarrow}^{+}+\delta m^{*}\psi_{\downarrow}\psi_{\uparrow})$,
where $\delta n_{\sigma}\equiv\delta n_{\sigma}\left({\bf r},t\right)$
and $\delta m\equiv\delta m\left({\bf r},t\right)$ are the normal
and anomalous density fluctuations, respectively, which must be determined
self-consistently \cite{minguzzi,bruun,LiuPRA2004,stringariPRL2009}.
In the linear approximation, the self-generated potential $\delta U$
plays the same role as the perturbation field when we calculate the
dynamic response using a static BCS Hamiltonian as the reference system.
This leads to coupled equations for density fluctuations. The linear
response is characterized by a matrix consisting of all two-particle
response functions \cite{bruun}: \begin{equation}
\chi\equiv\left\{ \begin{array}{cccc}
\langle\langle\hat{n}_{\uparrow}\hat{n}_{\uparrow}\rangle\rangle & \langle\langle\hat{n}_{\uparrow}\hat{n}_{\downarrow}\rangle\rangle & \langle\langle\hat{n}_{\uparrow}\hat{m}\rangle\rangle & \langle\langle\hat{n}_{\uparrow}\hat{m}^{+}\rangle\rangle\\
\langle\langle\hat{n}_{\downarrow}\hat{n}_{\uparrow}\rangle\rangle & \langle\langle\hat{n}_{\downarrow}\hat{n}_{\downarrow}\rangle\rangle & \langle\langle\hat{n}_{\downarrow}\hat{m}\rangle\rangle & \langle\langle\hat{n}_{\downarrow}\hat{m}^{+}\rangle\rangle\\
\langle\langle\hat{m}\hat{n}_{\uparrow}\rangle\rangle & \langle\langle\hat{m}\hat{n}_{\downarrow}\rangle\rangle & \langle\langle\hat{m}\hat{m}\rangle\rangle & \langle\langle\hat{m}\hat{m}^{+}\rangle\rangle\\
\langle\langle\hat{m}^{+}\hat{n}_{\uparrow}\rangle\rangle & \langle\langle\hat{m}^{+}\hat{n}_{\downarrow}\rangle\rangle & \langle\langle\hat{m}^{+}\hat{m}\rangle\rangle & \langle\langle\hat{m}^{+}\hat{m}^{+}\rangle\rangle\end{array}\right\} ,\end{equation}
 where $\langle\langle\hat{A}\hat{B}\rangle\rangle$ is the Fourier
transform of the retarded function $-i\Theta\left(t-t^{\prime}\right)\langle[\hat{A}({\bf r},t),\hat{B}({\bf r}^{\prime},t^{\prime})]\rangle$.
For simplicity, we abbreviate $\chi_{\sigma\sigma^{\prime}}\equiv\langle\langle\hat{n}_{\sigma}\hat{n}_{\sigma^{\prime}}\rangle\rangle$,
$\chi_{\sigma m}\equiv\langle\langle\hat{n}_{\sigma}\hat{m}\rangle\rangle$,
$\chi_{\sigma\bar{m}}\equiv\langle\langle\hat{n}_{\sigma}\hat{m}^{+}\rangle\rangle$,
$\chi_{m\bar{m}}\equiv\langle\langle\hat{m}\hat{m}^{+}\rangle\rangle$,
and so on. By solving the coupled equations for density fluctuations,
the standard RPA response function $\chi$ can be expressed in terms
of the static BCS response function $\chi^{0}$ \cite{bruun}, \begin{equation}
\chi=\chi^{0}\left[\hat{1}-U_{0}\chi^{0}{\cal G}\right]^{-1},\label{RPA}\end{equation}
 where ${\cal G}=\delta({\bf r}-{\bf r}^{\prime})[\sigma_{0}\otimes\sigma_{x}]$
is a direct product of two Pauli matrices $\sigma_{0}$ and $\sigma_{x}$
and the unit matrix $\hat{1}=\delta({\bf r}-{\bf r}^{\prime})[\sigma_{0}\otimes\sigma_{0}]$.
The dynamic structure factor $S_{\sigma\sigma^{\prime}}(\omega)$
is then obtained by using the fluctuation-dissipation theorem Eq.
(\ref{fluct-dissi-theorem}). In the weak-coupling regime, Eq.~(\ref{RPA})
has been be solved by calculating $\chi^{0}$ for a thermal average
of BCS quasiparticles \cite{minguzzi,bruun}.

\subsection{Strong-coupling RPA theory}

Here, we extend the RPA to the strongly interacting regime with an
arbitrarily large scattering length $a$, by properly renormalizing
the bare interaction strength $U_{0}$ and the two response functions
$\chi_{m\bar{m}}^{0}$ and $\chi_{\bar{m}m}^{0}$, which was found
to be suitable at the BEC-BCS crossover \cite{ourRPA}. The ultraviolet
divergence of these two functions is canceled exactly by the small
value of $U_{0}$, when the momentum cut-off goes to infinity. In
homogeneous systems, a careful account of the divergent terms in the
inverted matrix of the RPA equation (\ref{RPA}) leads to a concise
expression for the response functions: \begin{eqnarray}
\chi_{\uparrow\uparrow} & = & \chi_{\uparrow\uparrow}^{0}-\left[2\chi_{\uparrow\downarrow}^{0}\chi_{\uparrow m}^{0}\chi_{\uparrow\bar{m}}^{0}+\left(\chi_{\uparrow m}^{0}\right)^{2}\tilde{\chi}_{m\bar{m}}^{0}\right.\nonumber \\
 &  & \left.+\left(\chi_{\uparrow\bar{m}}^{0}\right)^{2}\tilde{\chi}_{\bar{m}m}^{0}\right]/\left[\tilde{\chi}_{m\bar{m}}^{0}\tilde{\chi}_{\bar{m}m}^{0}-\left(\chi_{\uparrow\downarrow}^{0}\right)^{2}\right],\label{RPAupup}\end{eqnarray}
 and \begin{equation}
\chi_{\uparrow\downarrow}=\chi_{\uparrow\uparrow}-\chi_{\uparrow\uparrow}^{0}+\chi_{\uparrow\downarrow}^{0},\label{RPAupdw}\end{equation}
 where the response functions with a tilde, i.e., $\tilde{\chi}_{m\bar{m}}^{0}\equiv\chi_{m\bar{m}}^{0}+\sum_{{\bf k}}m/(\hbar^{2}{\bf k}^{2})-m/(4\pi\hbar^{2}a)$,
become free from any ultraviolet divergence. Note that, we use a Leggett-BCS
ground state without inclusion of the Hartree-Fock term in the quasiparticle
spectrum. Therefore, in the BCS regime our treatment does not account
for the leading interaction effect. At the crossover, however, it
does capture the dominant pairing gap. Note also that, the RPA method
accounts for single particle-hole excitations. Higher correlations
such as multi-particle-hole excitations are neglected.

In the presence of a harmonic trap, the renormalization procedure
becomes cumbersome because of the discrete energy levels. It is convenient
to use a local density approximation (LDA) \cite{LDABF,LDAFG}, which
treats the system as a collection of many homogeneous cells with local
chemical potential, $\mu({\bf r})=\mu-V_{T}({\bf r})$, where $V_{T}({\bf r})=M(\omega_{x}^{2}x^{2}+\omega_{y}^{2}y^{2}+\omega_{z}^{2}z^{2}{\bf )}/2$
is the harmonic trapping potential. The LDA treatment is valid for
a large number of atoms such as $N\sim10^{5}$ as in experiments.
At a given temperature and scattering length, we solve the Leggett-BCS
equation with local chemical potential for the local pairing gap and
calculate the static response function $\chi^{0}$, then solve the
local RPA density response functions using Eqs. (\ref{RPAupup}) and
(\ref{RPAupdw}), and finally obtain the total RPA responses by integrating
over the whole trap. To calculate the dimensionless interaction strength
$1/(k_{F}a)$, we use the Fermi wave-vector $k_{F}=\sqrt{2mE_{F}/\hbar^{2}}$
at the trap center, where the trapped Fermi energy is given by $E_{F}=\hbar(3N\omega_{x}\omega_{y}\omega_{z})^{1/3}$
\cite{GiorginiRMP}.

\subsection{Comparison of theory with experiment at zero temperature}

Fig. 3 shows the zero-temperature spin parallel, anti-parallel, and
total DSF at a transferred wave-vector $q=5k_{F}$ in the BEC-BCS
crossover, calculated using the above RPA procedure for a trapped
Fermi gas \cite{ourRPA}. In addition to a broad response at the atomic
resonance frequency $\omega_{R}=\hbar q^{2}/(2M)=25E_{F}/\hbar$ caused
by resonant scattering of atoms, a much narrower peak develops at
about $\omega_{R}/2$ with increased coupling. This is precisely what
has been observed in the Bragg spectroscopy experiment. Physically,
it is the Bogoliubov-Anderson phonon mode of a Fermi superfluid at
large wave-vectors, which evolves continuously into a Bogoliubov mode
of molecules towards the BEC limit \cite{combescot}. The molecular
peak is mostly evident in $S_{\uparrow\downarrow}$ as there is no
background atomic response.

\begin{figure}[htp]

\begin{centering}
\includegraphics[clip,width=0.45\textwidth]{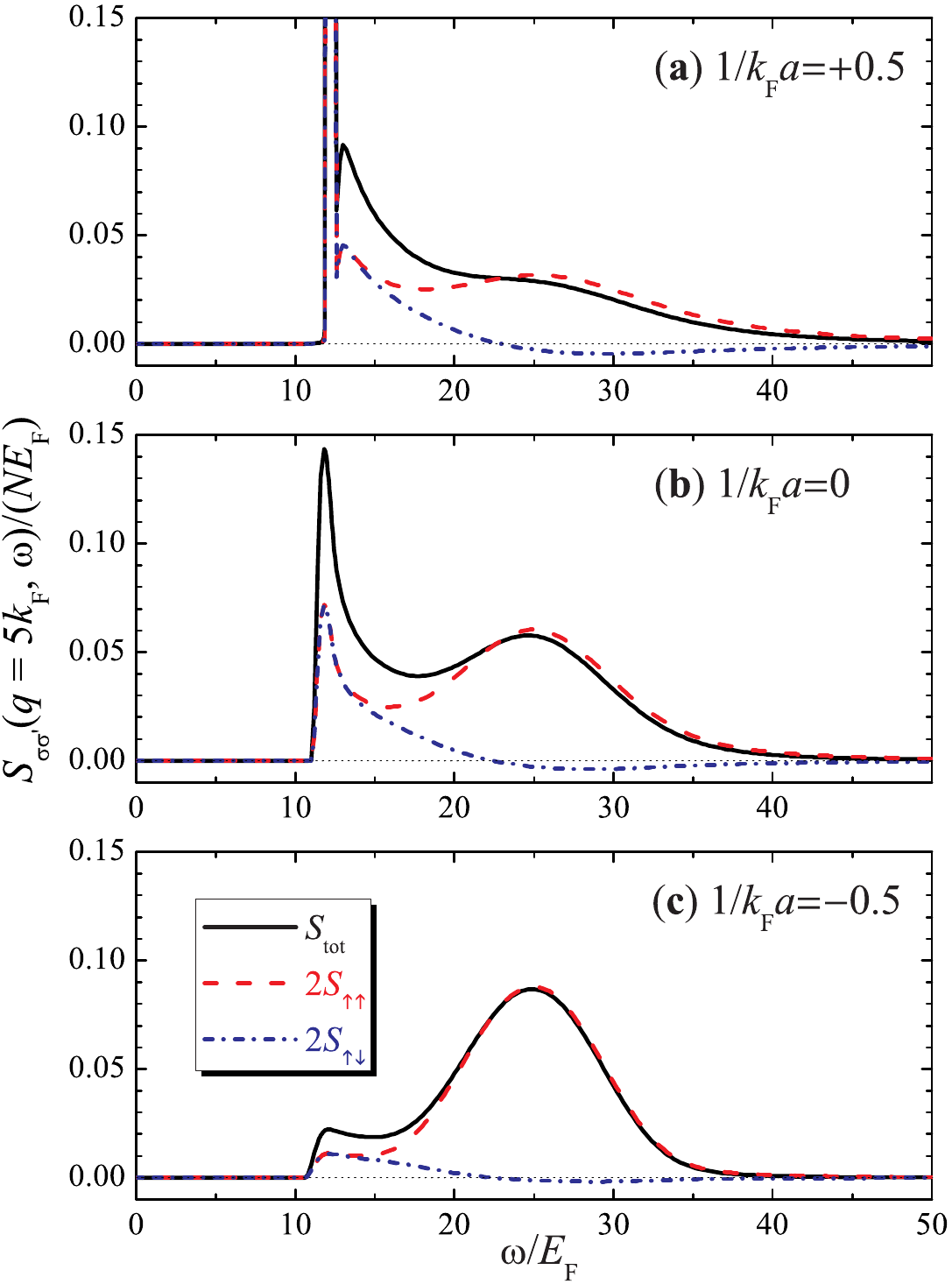} 
\par\end{centering}

\caption{(color online) Zero temperature spin parallel $S_{\uparrow\uparrow}({\bf q},\omega)$
(dashed lines), anti-parallel $S_{\uparrow\downarrow}({\bf q},\omega)$
(dot-dashed lines), and total dynamic structure factor $S({\bf q},\omega)=2[S_{\uparrow\uparrow}+S_{\uparrow\downarrow}]$
(solid lines) across the BEC-BCS crossover: $1/k_{F}a=0.5$ (a), $0.0$
(b), and $-0.5$ (c). The negative weight in $S_{\uparrow\downarrow}$
at about the atomic resonance frequency is consistent with the exact
sum rule $\int\omega S_{\uparrow\downarrow}({\bf q},\omega)d\omega=0$.
The figure is reproduced from ref. \cite{ourRPA} with permission.}

\label{fig3} 
\end{figure}

\begin{figure}[htp]

\begin{centering}
\includegraphics[clip,width=0.5\textwidth]{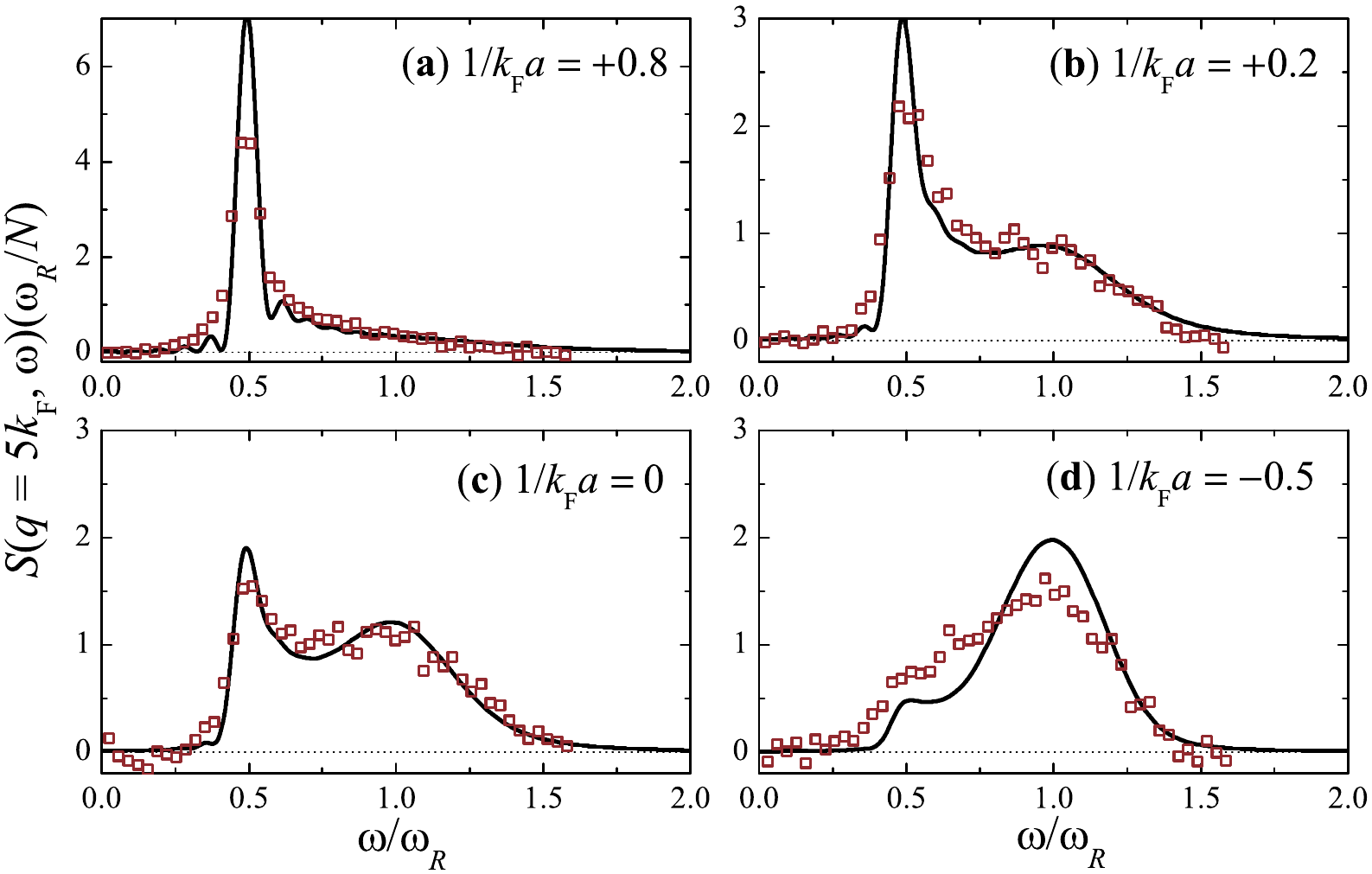} 
\par\end{centering}

\caption{(color online) Quantitative comparison of theoretical and experimental
Bragg spectra. The RPA prediction (lines) agrees well with the experimental
data (empty squares) \cite{BraggSwin} at the BEC-BCS crossover, with
no free parameters. The figure is copied from ref. \cite{ourRPA}
with permission.}

\label{fig4} 
\end{figure}

To make a quantitative comparison with the experimental spectra, we
take into account the spectral broadening due to the finite experimental
Bragg pulse duration ($\tau_{Br}=40$ $\mu$s) \cite{BraggSwin}.
We calculate theoretically, \begin{equation}
S_{expt}\left({\bf q},\omega\right)\propto\frac{1}{\pi\sigma}\int_{-\infty}^{\infty}d\omega^{\prime}S({\bf q},\omega^{\prime}){\rm sinc^{2}}\left[\frac{\omega-\omega^{\prime}}{\sigma}\right],\label{Pimp}\end{equation}
 where ${\rm sinc}(x)=\sin(x)/x$ and the energy resolution $\sigma=2/\tau_{Br}$.
We find $\sigma\approx0.68E_{F}$ $\approx0.027\omega_{R}$. Fig.
4 presents a comparison of the experimental data (open squares) with
the RPA predictions (lines) for the Bragg spectra. Without any free
parameters, our RPA predictions agree well with the experimental results
in the unitarity regime ($1/k_{F}a=0.0$ and $0.2$) and BEC regime
($1/k_{F}a=0.8$). The agreement on the BCS side ($1/k_{F}a=-0.5$),
however, becomes a bit worse. The quantitative agreement around unitarity
is very compelling, since the RPA was assumed to be unreliable in
the (strongly interacting) regime of large pair fluctuations. Our
comparison indicates that the RPA is able to describe the dynamical
properties of the BEC-BCS crossover, at least at zero temperature
and large momenta. High order multi-particle-hole excitations, absent
in the RPA theory, seems to be negligibly small at large momenta.
The somewhat poorer agreement at $1/k_{F}a=-0.5$ can be attributed
to the mean-field shift, which is ignored in the RPA but dominates
for sufficiently weak interactions.

\begin{figure}[htp]

\begin{centering}
\includegraphics[clip,width=0.45\textwidth]{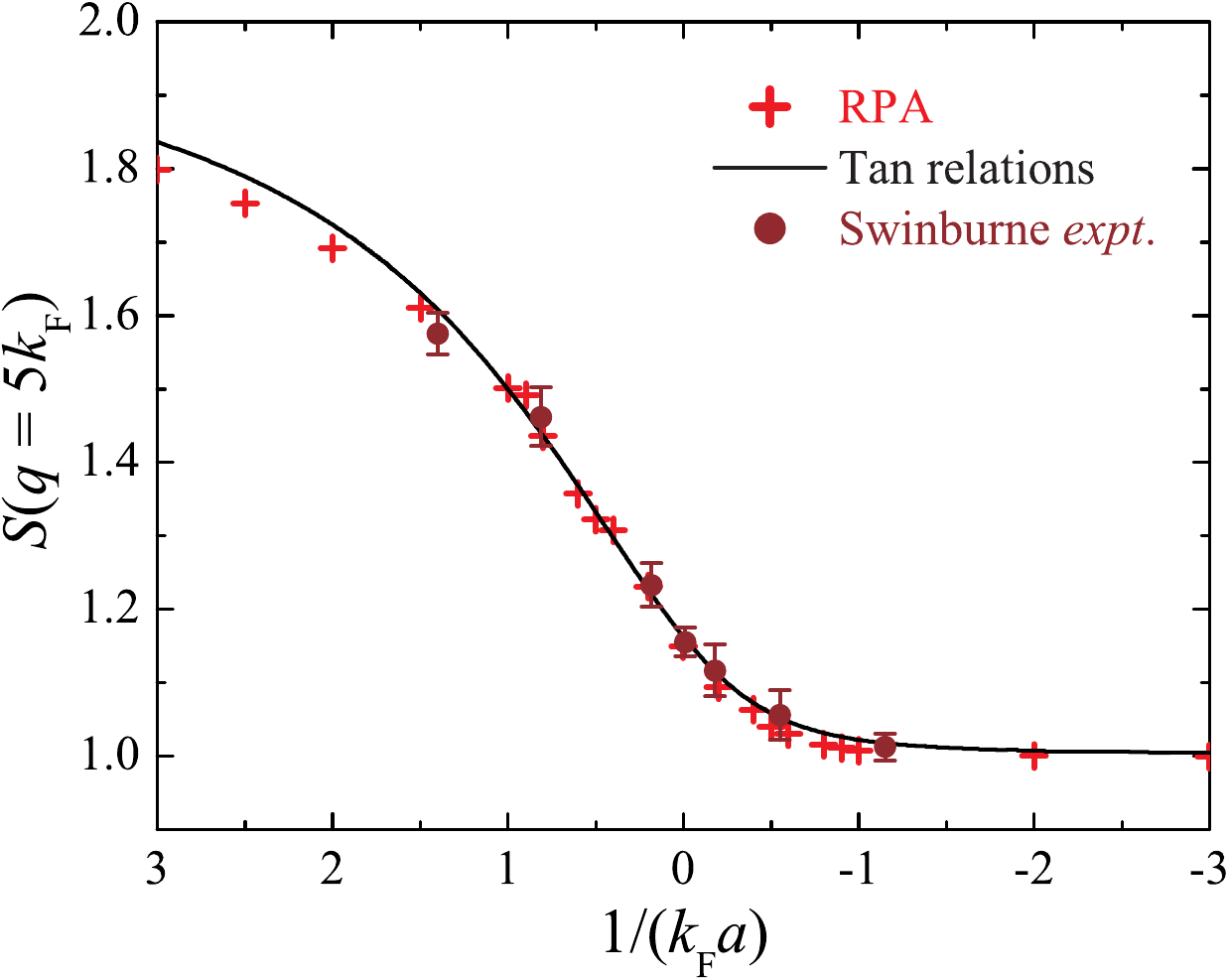} 
\par\end{centering}

\caption{(color online) Quantitative comparison between theory and experiment
for the zero temperature static structure factor across the crossover.
With no free parameters our RPA prediction (plus symbols) agrees well
with the experimental data for $S({\bf q})$ (solid circles with error
bars) \cite{TanSwin1} and an independent theoretical result based
on the exact Tan relations (solid line). The figure is reproduced
from ref. \cite{ourRPA} with permission.}

\label{fig5} 
\end{figure}

The good agreement between the RPA theory and the Bragg experiment
is further confirmed by comparing the SSF at zero temperature, as
reported in Fig. 5. Note that, the measured SSF is not affected by
the spectral broadening due to the finite Bragg pulse duration. It
is evident that the RPA prediction (plus symbols) fits very well with
the experimental data (solid circles). We show also an independent
theoretical prediction from exact Tan relations, which will be discussed
in detail in Sec. V. The two theoretical predictions agree well with
each other.

\section{Quantum cluster expansion}

Let us now consider the non-perturbative quantum virial expansion
\cite{ourVirialDSF}. Virial expansion has proved to be an efficient
method for studying the high-temperature properties of ultracold atomic
Fermi gases \cite{HLDEoS,hovirial,lhdprl2009,lhdpra2010,lhdprb2010,veakw,vecontact}.
This method utilizes the fact that in the high temperature limit,
the chemical potential $\mu$ diverges to $-\infty$ and the fugacity
$z\equiv\exp(\mu/k_{B}T)\ll1$ is a well-defined small expansion parameter.
Thus, one may expand any physical quantities of interest in powers
of fugacity, no matter how strong the interactions.

\subsection{Virial expansion of DSF}

We construct first the virial expansion for the dynamic susceptibility
$\chi_{\sigma\sigma^{\prime}}\left({\bf r},{\bf r}^{\prime};\tau>0\right)$,
which is given by, \begin{equation}
\chi_{\sigma\sigma^{\prime}}\equiv-\frac{\text{Tr}\left[e^{-\beta\left({\cal H}-\mu{\cal N}\right)}e^{{\cal H}\tau}\hat{n}_{\sigma}\left({\bf r}\right)e^{-{\cal H}\tau}\hat{n}_{\sigma^{\prime}}\left({\bf r}^{\prime}\right)\right]}{\text{Tr}e^{-\beta\left({\cal H}-\mu{\cal N}\right)}}.\end{equation}
 At high temperatures, Taylor-expanding in terms of the powers of
small fugacity $z\equiv\exp(\mu/k_{B}T)\ll1$ leads to $\chi_{\sigma\sigma^{\prime}}\left({\bf r},{\bf r}^{\prime};\tau\right)=(zX_{1}+z^{2}X_{2}+\cdots)/(1+zQ_{1}+z^{2}Q_{2}+\cdots)=zX_{1}+z^{2}\left(X_{2}-X_{1}Q_{1}\right)+\cdots$,
where we have introduced the cluster functions $X_{n}=-$ Tr$_{n}[e^{-\beta{\cal H}}e^{\tau{\cal H}}\hat{n}_{\sigma}({\bf r)}e^{-\tau{\cal H}}\hat{n}_{\sigma^{\prime}}({\bf r}^{\prime})]$
and $Q_{n}=$Tr$_{n}[e^{-\beta{\cal H}}]$, with $n$ denoting the
number of particles in the cluster and Tr$_{n}$ denoting the trace
over $n$-particle states of proper symmetry. We shall refer to the
above expansion as the virial expansion of dynamic susceptibilities,
$\chi_{\sigma\sigma^{\prime}}\left({\bf r},{\bf r}^{\prime};\tau\right)=z\chi_{\sigma\sigma^{\prime},1}\left({\bf r},{\bf r}^{\prime};\tau\right)+z^{2}\chi_{\sigma\sigma^{\prime},2}\left({\bf r},{\bf r}^{\prime};\tau\right)+\cdots,$
where, \begin{eqnarray}
\chi_{\sigma\sigma^{\prime},1}\left({\bf r},{\bf r}^{\prime};\tau\right) & = & X_{1},\nonumber \\
\chi_{\sigma\sigma^{\prime},2}\left({\bf r},{\bf r}^{\prime};\tau\right) & = & X_{2}-X_{1}Q_{1},\ \text{etc}.\end{eqnarray}
 Accordingly, we shall write for the dynamic structure factors, \begin{equation}
S_{\sigma\sigma^{\prime}}\left({\bf q},\omega\right)=zS_{\sigma\sigma^{\prime},1}\left({\bf q},\omega\right)+z^{2}S_{\sigma\sigma^{\prime},2}\left({\bf q},\omega\right)+\cdots.\end{equation}

The calculation of the $n$-th expansion coefficient requires the
knowledge of all solutions up to $n$-body, including both the eigenvalues
and eigenstates \cite{lhdprl2009,lhdpra2010}. Here we aim to calculate
the leading effect of interactions, which contribute to the 2nd-order
expansion function \cite{ourVirialDSF}. For this purpose, it is convenient
to define $\Delta\chi_{\sigma\sigma^{\prime},2}\equiv\left\{ \chi_{\sigma\sigma^{\prime},2}\right\} ^{(I)}=\left\{ X_{2}\right\} ^{(I)}$
and $\Delta S_{\sigma\sigma^{\prime},2}\equiv\left\{ S_{\sigma\sigma^{\prime},2}\right\} ^{(I)}$.
The notation $\left\{ {}\right\} ^{(I)}$ means the contribution due
to interactions inside the bracketed term, so that $\left\{ X_{2}\right\} ^{(I)}=X_{2}-X_{2}^{(1)}$,
where the superscript {}``1'' in $X_{2}^{(1)}$ denotes quantities
for a noninteracting system. We note that the inclusion of the 3rd-order
expansion function is straightforward, though involving more numerical
effort.

It is easy to see that $\Delta\chi_{\sigma\sigma^{\prime},1}=0$,
according to the definition of notation $\left\{ {}\right\} ^{(I)}$.
To calculate the 2nd-order expansion function for the dynamic susceptibility,
$\Delta\chi_{\sigma\sigma^{\prime},2}=-\left\{ Tr_{\uparrow\downarrow}\left[e^{-\beta{\cal H}}e^{\tau{\cal H}}\hat{n}_{\sigma}\left({\bf r}\right)e^{-\tau{\cal H}}\hat{n}_{\sigma^{\prime}}\left({\bf r}^{\prime}\right)\right]\right\} ^{(I)}$,
we insert the identity $\sum_{Q}\left|Q\right\rangle \left\langle Q\right|={\bf \hat{1}}$
and take the trace over the state $P$, i.e., $\Delta\chi_{\sigma\sigma^{\prime},2}=-\sum_{P,Q}\left\{ e^{-\beta E_{P}+\tau(E_{P}-E_{Q})}\left\langle P\left|\hat{n}_{\sigma}\right|Q\right\rangle \left\langle Q\left|\hat{n}_{\sigma^{\prime}}\right|P\right\rangle \right\} ^{(I)}$.
Here, $P$ and $Q$ are the two-atom eigenstates with energies $E_{P}$
and $E_{Q}$, respectively. Expressing the density operator in the
first quantization: $\hat{n}_{\uparrow}\left({\bf r}\right)=\sum_{i}\delta\left({\bf r}-{\bf r}_{i\uparrow}\right)$
and $\hat{n}_{\downarrow}\left({\bf r}\right)=\sum_{j}\delta({\bf r}-{\bf r}_{j\downarrow})$,
it is straightforward to show that, \begin{equation}
\Delta\chi_{\sigma\sigma^{\prime},2}=-\sum_{P,Q}\left\{ e^{-\beta E_{P}+\tau\left(E_{P}-E_{Q}\right)}C_{\sigma\sigma^{\prime}}^{PQ}\left({\bf r},{\bf r}^{\prime}\right)\right\} ^{(I)},\end{equation}
 where \begin{equation}
C_{\uparrow\uparrow}^{PQ}\equiv\int d{\bf r}_{2}d{\bf r}_{2}^{\prime}\left[\Phi_{P}^{*}\Phi_{Q}\right]\left({\bf r},{\bf r}_{2}\right)[\Phi_{Q}^{*}\Phi_{P}]\left({\bf r}^{\prime},{\bf r}_{2}^{\prime}\right)\end{equation}
 and \begin{equation}
C_{\uparrow\downarrow}^{PQ}\equiv\int d{\bf r}_{1}d{\bf r}_{2}[\Phi_{P}^{*}\Phi_{Q}]\left({\bf r},{\bf r}_{2}\right)[\Phi_{Q}^{*}\Phi_{P}]\left({\bf r}_{1},{\bf r}^{\prime}\right).\end{equation}
 The dynamic structure factor can be obtained by taking the analytic
continuation. This result is $\Delta S_{\sigma\sigma^{\prime},2}\left({\bf r},{\bf r}^{\prime};\omega\right)=\sum_{P,Q}\left\{ \delta\left(\omega+E_{P}-E_{Q}\right)e^{-\beta E_{P}}C_{\sigma\sigma^{\prime}}^{PQ}\left({\bf r},{\bf r}^{\prime}\right)\right\} ^{(I)}$.
Applying a further Fourier transform with respect to ${\bf x}={\bf r}-{\bf r}^{\prime}$
and integrating over ${\bf R}=({\bf r}+{\bf r}^{\prime})/2$, we obtain
the response $\Delta S_{\sigma\sigma^{\prime},2}\left({\bf q},\omega\right)$,
\begin{equation}
\Delta S_{\sigma\sigma^{\prime},2}=\sum_{P,Q}\left\{ \delta\left(\omega+E_{P}-E_{Q}\right)e^{-\beta E_{P}}F_{\sigma\sigma^{\prime}}^{PQ}\left({\bf q}\right)\right\} ^{(I)},\end{equation}
 where $F_{\sigma\sigma^{\prime}}^{PQ}\left({\bf q}\right)=\int d{\bf r}d{\bf r}^{\prime}e^{-i{\bf q\cdot}({\bf r}-{\bf r}^{\prime})}C_{\sigma\sigma^{\prime}}^{PQ}\left({\bf r},{\bf r}^{\prime}\right)$. 

The calculation of $C_{\sigma\sigma^{\prime}}^{PQ}\left({\bf r},{\bf r}^{\prime}\right)$
or $F_{\sigma\sigma^{\prime}}^{PQ}\left({\bf q}\right)$ is straightforward
but tedious, by using the two-atom solutions in an isotropic harmonic
trap $m\omega_{0}^{2}r^{2}/2$. We refer to ref. \cite{ourVirialDSF}
for details. The final result is given by, \begin{equation}
\Delta S_{\sigma\sigma^{\prime},2}=B\sqrt{\frac{m}{\pi}}\sum_{p2q2}\left\{ e^{-\frac{\beta\left(\tilde{\omega}+\epsilon_{p2}-\epsilon_{q2}\right)^{2}}{2\omega_{R}}}e^{-\beta\epsilon_{p2}}A_{p2q2}^{\sigma\sigma^{\prime}}\right\} ^{(I)}\,\,,\label{ddsf2}\end{equation}
 where $B\equiv(k_{B}T)^{5/2}/(q\hbar^{4}\omega_{0}^{3})$, $\tilde{\omega}=\omega-\omega_{R}/2$,
and \begin{eqnarray}
A_{p2q2}^{\sigma\sigma^{\prime}} & = & (-1)^{l(1-\delta_{\sigma\sigma^{\prime}})}(2l+1)\times\nonumber \\
 &  & \left[\int_{0}^{\infty}dxx^{2}j_{l}\left(\frac{qx}{2}\right)\phi_{n_{p}l_{p}}\left(x\right)\phi_{n_{q}l_{q}}\left(x\right)\right]^{2}.\label{Arel}\end{eqnarray}
 In Eq. (\ref{Arel}), we specify $p2=\{n_{p}l_{p}\}$ and $q2=\{n_{q}l_{q}\}$,
and $l=\max\{l_{p},l_{q}\}$ for the two-atom relative radial wave
functions $\phi$ with energy $\epsilon$. We require that either
$l_{p}$ or $l_{q}$ should be zero (i.e., $\min\{l_{p},l_{q}\}=0$),
otherwise $A_{p2q2}^{\sigma\sigma^{\prime}}$ will be cancelled exactly
by the non-interacting terms.

Together with the non-interacting DSF $S_{\sigma\sigma^{\prime}}^{\left(1\right)}$,
we calculate directly the interacting structure factor, \begin{equation}
S_{\sigma\sigma^{\prime}}({\bf q},\omega)=S_{\sigma\sigma^{\prime}}^{\left(1\right)}({\bf q},\omega)+z^{2}\Delta S_{\sigma\sigma^{\prime},2},\end{equation}
 once the fugacity $z$ is determined by the virial expansion for
equation of state.

\subsection{Comparison of theory with experiment at high temperatures}

Considerable insight into the dynamic structure factor of a strongly
correlated Fermi gas can already be seen from Eq. (\ref{ddsf2}),
in which the spectrum is peaked roughly at $\omega_{R,mol}=\omega_{R}/2$,
the resonant frequency for molecules. Therefore, the peak is related
to the response of molecules with mass $M=2m$. Eq. (\ref{ddsf2})
shows clearly how the molecular response develops with the modified
two-fermion energies and wave functions as the interaction strength
increases. In the BCS limit where $\Delta S_{\sigma\sigma^{\prime},2}$
is small, the response is determined by the non-interacting background
$S_{\sigma\sigma^{\prime}}^{\left(1\right)}$ that peaks at $\omega_{R}$.
In the extreme BEC limit ($a\rightarrow0^{+}$), however, $\Delta S_{\sigma\sigma^{\prime},2}$
dominates. The sum in $\Delta S_{\sigma\sigma^{\prime},2}$ is exhausted
by the (lowest) tightly bound state $\phi_{rel}(x)\simeq\sqrt{2/a}e^{-x/a}$
with energy $\epsilon_{rel}\simeq-\epsilon_{B}\equiv-\hbar^{2}/(ma^{2})$.
The chemical potential of molecules is given by $\mu_{m}=2\mu+\epsilon_{B}$.
Therefore, the DSF of fermions takes the form, \begin{equation}
S_{\sigma\sigma^{\prime}}^{BEC}\simeq z_{m}B\sqrt{\frac{M}{\pi}}\exp\left[-\frac{\beta(\omega-\omega_{R,mol})^{2}}{4\omega_{R,mol}}\right],\label{dsfbec}\end{equation}
 where $z_{m}=e^{\beta\mu_{m}}$ is the molecular fugacity. This peaks
at the molecular resonant energy. As anticipated, Eq. (\ref{dsfbec})
is exactly the leading virial expansion term in the DSF of non-interacting
molecules. It is clear that $S_{\uparrow\uparrow}({\bf q},\omega)\simeq S_{\uparrow\downarrow}({\bf q},\omega)$
in the BEC limit, since the spin structure in a single molecule can
no longer be resolved.

\begin{figure}[htp]

\begin{centering}
\includegraphics[clip,width=0.45\textwidth]{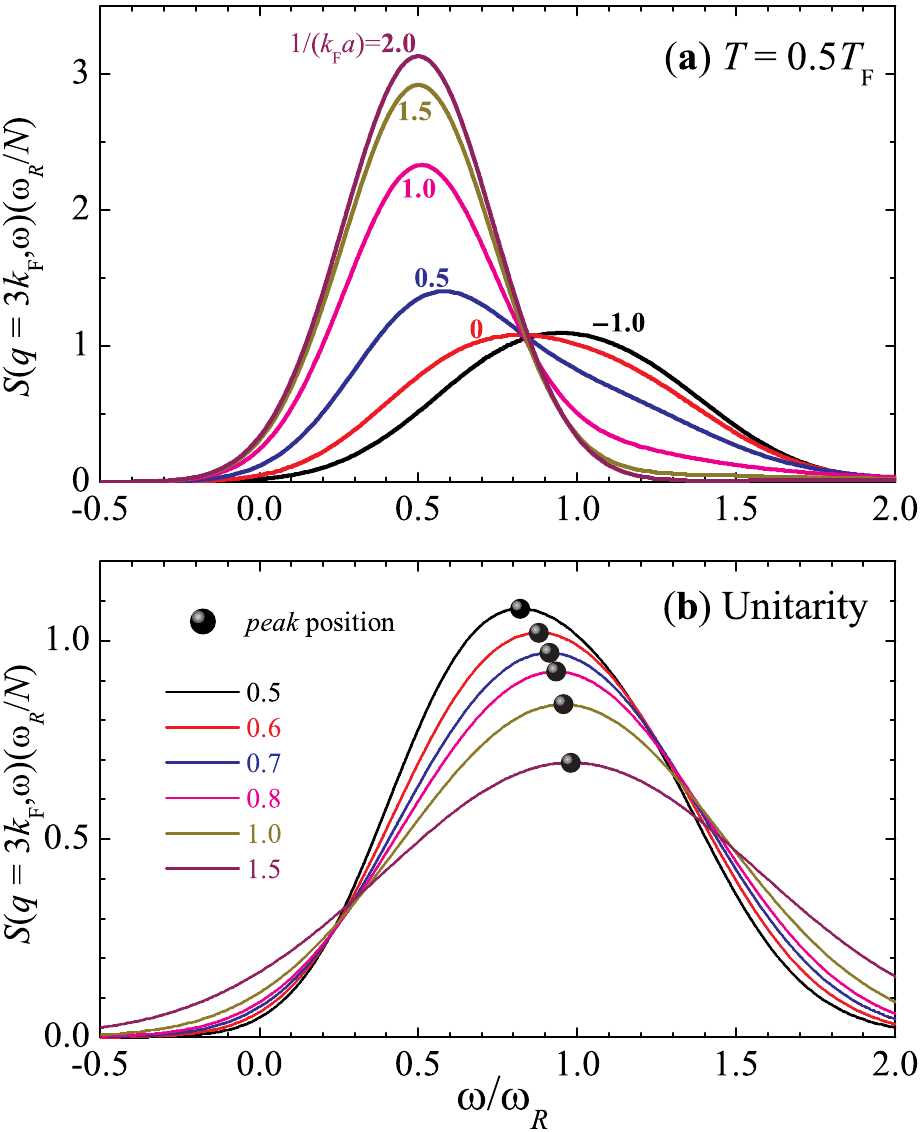} 
\par\end{centering}

\caption{(color online) (a) Evolution of dynamic structure factor of a trapped
Fermi gas in the BEC-BCS crossover with increasing interaction strength
$1/(k_{F}a)$ at $T=0.5T_{F}$. (b) Temperature dependence of dynamic
structure factor of a trapped unitary Fermi gas. The dark circles
indicate the peak position of spectra. The transferred wave-vector
is $q=3k_{F}$. The figure is reproduced from ref. \cite{ourVirialDSF}
with permission.}

\label{fig6} 
\end{figure}

To understand the intermediate regime, in Fig. 6a we report numerical
results for the DSF as the interaction strength increases from the
BCS to BEC regimes at $T=0.5T_{F}$ \cite{ourVirialDSF}. The temperature
dependence of the DSF in the unitary limit is shown in Fig. 6b \cite{ourVirialDSF}.
In a trapped gas with total number of fermions $N$, we use the zero
temperature Thomas-Fermi wave vector $k_{F}=(24N)^{1/6}/a_{ho}$ and
temperature $T_{F}=(3N)^{1/3}\hbar\omega_{0}/k_{B}$ as characteristic
units. In accord with the experiment \cite{BraggSwin,TanSwin2}, we
take a large transferred momentum of $q=3k_{F}$. At $T=0.5T_{F}$,
A smooth transition from atomic to molecular responses is evident
as the interaction parameter $1/(k_{F}a)$ increases, in qualitative
agreement with the experimental observation (c.f. Fig. 1). In the
unitary limit, the peak of total DSF shifts towards the molecular
recoil frequency, as indicated by the dark circles. This red-shift
is again in qualitative agreement with experiment (c.f. Fig. 2).

\begin{figure}[htp]

\begin{centering}
\includegraphics[clip,width=0.45\textwidth]{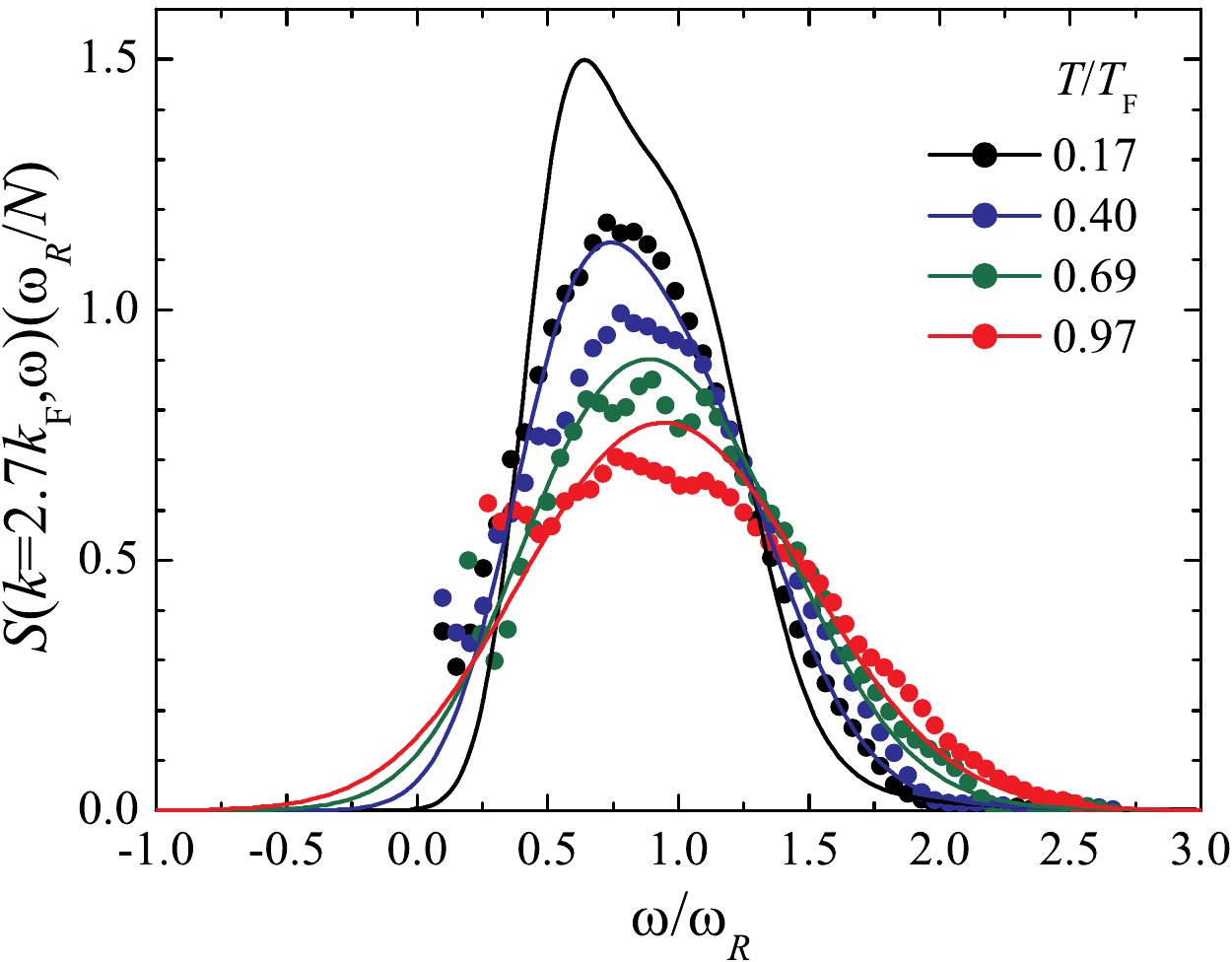} 
\par\end{centering}

\caption{(color online) Comparison between theory and experiment for the dynamic
structure factor of a trapped unitary Fermi gas at finite temperatures.
Here, the transferred wave-vector is $q=2.7k_{F}$. The figure is
reproduced from ref. \cite{TanSwin3} with permission.}

\label{fig7} 
\end{figure}

For a close comparison, we plot in Fig. 7 the virial expansion prediction
and experimental data for the DSF at several temperatures in the unitary
limit. The theory is in very good agreement with experimental data
at high temperatures (i.e., $T=0.97T_{F}$ and $T=0.69T_{F}$) \cite{TanSwin2,TanSwin3},
where the fugacity $z$ is less than $1$. Towards low temperatures,
the agreement becomes worse. However, the virial expansion does capture
the qualitative feature of the DSF, for temperature down to the superfluid
transition temperature $T_{c}\sim0.2T_{F}$.

\section{Exact Tan relations}

In this section, we present an exact relation for the large-momentum
behavior of the spin-antiparallel static structure factor $S_{\uparrow\downarrow}(q)$,
showing that it has a simple universal power-law ($1/q$) tail \cite{ourTanSSF}.
The relation we consider, hereafter referred to as the structure-factor
Tan relation, belongs to the family of exact relations obtained by
Shina Tan in 2005 \cite{TanRelations}, which link the short-range,
large-momentum, and high-frequency asymptotic behavior of many-body
systems to their thermodynamic properties. For instance, the momentum
distribution \cite{TanRelations} and rf-spectroscopy \cite{rfsumrule}
fall off as $q^{-4}$ and $\omega^{-5/2}$, respectively. All the
Tan relations are related to each other by a single{\em \ }coefficient
${\cal I}$, referred to as the integrated contact density or contact.
The contact measures the probability of two fermions with unlike spins
being close together \cite{braaten2008}. It also links the short-range
behavior to thermodynamics via the adiabatic relation \cite{TanRelations},
$dE/d(-1/a)=\hbar^{2}{\cal I}/(4\pi m)$, which gives the change in
the total energy $E$ due to adiabatic changes in the scattering length.
The fundamental importance of the Tan relations arises from their
wide applicability. They are useful at both zero or finite temperature,
superfluid or normal phase, homogeneous or trapped, few-body or many-body
systems.

We consider also the large-frequency tail of the DSF at large momentum
\cite{universaldsf}. By using Feynman diagrams, we show that the
spin parallel and antiparallel DSFs have respectively a tail of the
form $\sim\pm\omega^{-5/2}$ for $\omega\rightarrow\infty$, decaying
slower than the total DSF found previously ($\sim\omega^{-7/2}$)
\cite{son,taylor}.

\subsection{The structure-factor Tan relation}

The structure factor Tan relation for $S_{\uparrow\downarrow}(q)$
follows directly from the short-range behavior of the pair correlation
function $g_{\uparrow\downarrow}({\bf r})\equiv\int d{\bf R}\left\langle \hat{\rho}_{\uparrow}\left({\bf R}-{\bf r}/2\right)\hat{\rho}_{\downarrow}\left({\bf R}+{\bf r}/2\right)\right\rangle $,
which diverges as \cite{TanRelations} \begin{equation}
g_{\uparrow\downarrow}({\bf r}\rightarrow0)\simeq\frac{{\cal I}}{16\pi^{2}}\left(\frac{1}{r^{2}}-\frac{2}{ar}\right).\end{equation}
 A Fourier transformation of $g_{\uparrow\downarrow}({\bf r}\rightarrow0)$
then leads to \cite{ourTanSSF} \begin{equation}
S_{\uparrow\downarrow}\left(q\gg k_{F}\right)\simeq\frac{{\cal I}}{4Nq}\left[1-\frac{4}{\pi aq}\right]\equiv\frac{{\cal I}}{Nk_{F}}t\left(q\right),\label{TanSSF}\end{equation}
 where $k_{F}$ is the Fermi wave-vector and $N$ is the total number
of atoms. On the right hand side of the above equation, we have defined
$t(q)\equiv[k_{F}/(4q)][1-4/(\pi aq)]$. Eq. (\ref{TanSSF}) holds
in a scaling region of sufficiently large $q$ near the unitarity
limit ($a\rightarrow\pm\infty$) so that the next-order correction
in the bracket ($\propto1/(aq)$) is small compared to the leading
term of $1$.

The power-law tail of $1/q$ in the structure-factor relation is more
amenable for experimental investigation than the $q^{-4}$ or $\omega^{-5/2}$
tail in the momentum distribution or in rf spectroscopy. The fast
decay due to the higher power law in these latter two cases imposes
very stringent signal-to-noise requirements for a given range of momentum
or frequency. As we already shown before, experimentally the static
structure factor $S(q)=S_{\uparrow\uparrow}(q)+S_{\uparrow\downarrow}(q)$
can be readily measured using two-photon Bragg spectroscopy on a balanced
two-component atomic Fermi gas near a Feshbach resonance. In the large-$q$
limit, to an good approximation $S_{\uparrow\uparrow}(q)\simeq1$.
One thus can directly determine the spin-antiparallel SSF $S_{\uparrow\downarrow}(q)=S(q)-1$
and verify the simple $1/q$ asymptotic behavior.

\begin{figure}[htp]

\begin{centering}
\includegraphics[clip,width=0.45\textwidth]{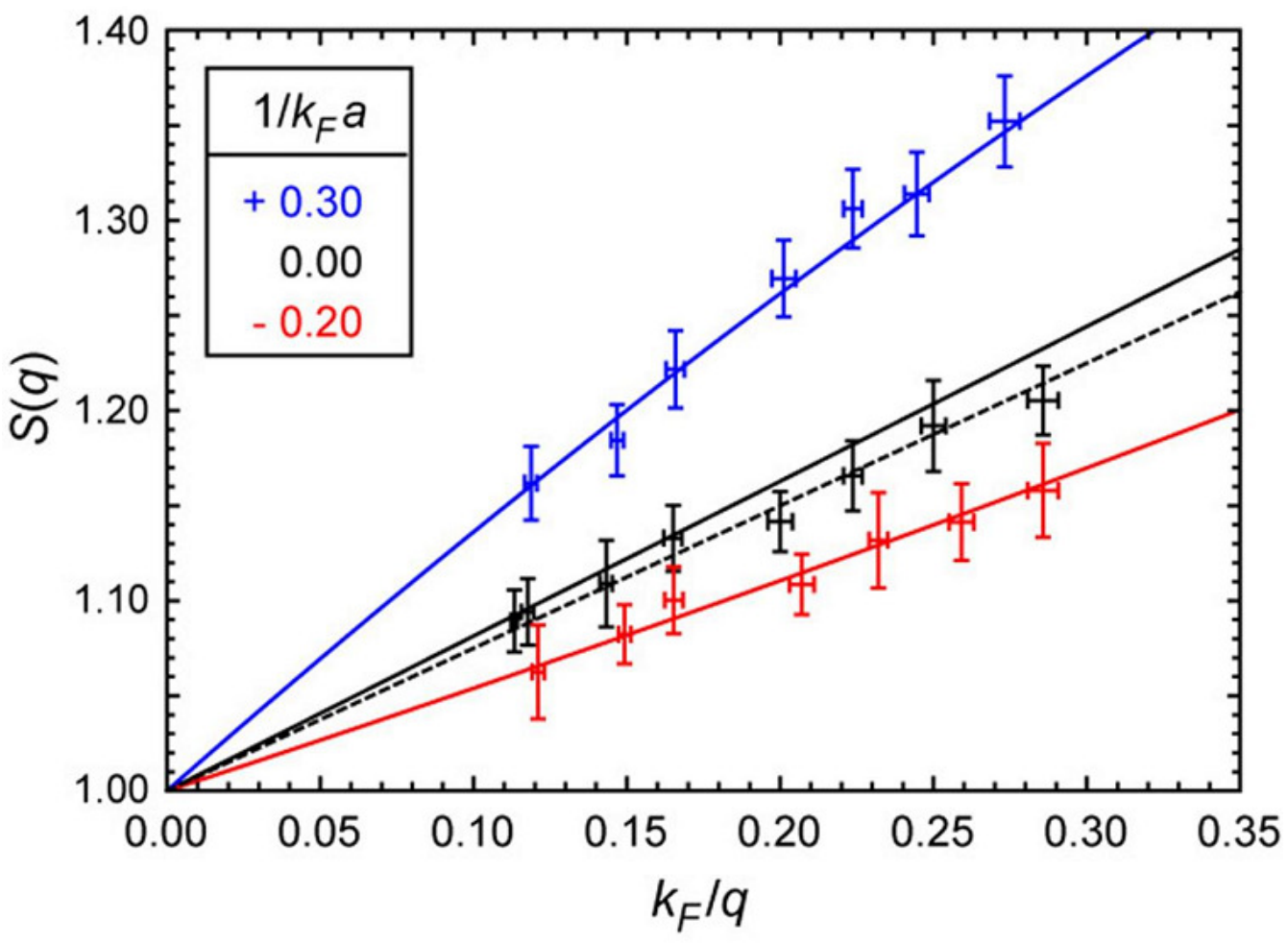} 
\par\end{centering}

\caption{(color online) Experimental confirmation of the structure-factor Tan
relation Eq. (\ref{TanSSF}). The measured static structure factor
of a trapped crossover Fermi gas at lowest attainable temperature
is plotted as a function of inverse transferred wave-vector. The solid
lines show the theoretical prediction of Eq. (\ref{TanSSF}). The
dashed line is a linear fit to the experimental data in the unitary
limit. It lies slightly below the predicted solid line, presumably
due to the finite temperature effect \cite{TanSwin1}. The figure
is copied from ref. \cite{TanSwin1} with permission.}

\label{fig8} 
\end{figure}

In Fig. 8, we show the experimental confirmation of Eq. (\ref{TanSSF})
at lowest experimental temperatures \cite{TanSwin1}. The SSF has
been measured as a function of the inverse transferred momentum $k_{F}/q$
at three interaction parameters, shown by the symbols with error bars
\cite{TanSwin1}. The solid lines plot the theoretical prediction
of Eq. (\ref{TanSSF}). The contact ${\cal I}$ used in the equation
has been calculated by using Tan's adiabatic relation and the reliable
theoretical results for zero-temperature equation of state \cite{vecontact}.
The agreement between theory and experiment is quantitative. At $1/(k_{F}a)=+0.3$,
the data depart from a straight line displaying the downward curvature
consistent with the first order term in Eq. (\ref{TanSSF}). A similar
upward curvature is seen at $1/(k_{F}a)=-0.2$. The simple structure-factor
Tan relation is seen to accurately describe the SSF in the BEC-BCS
crossover demonstrating the wide applicability of the Tan relations.

\begin{figure}[htp]

\begin{centering}
\includegraphics[clip,width=0.45\textwidth]{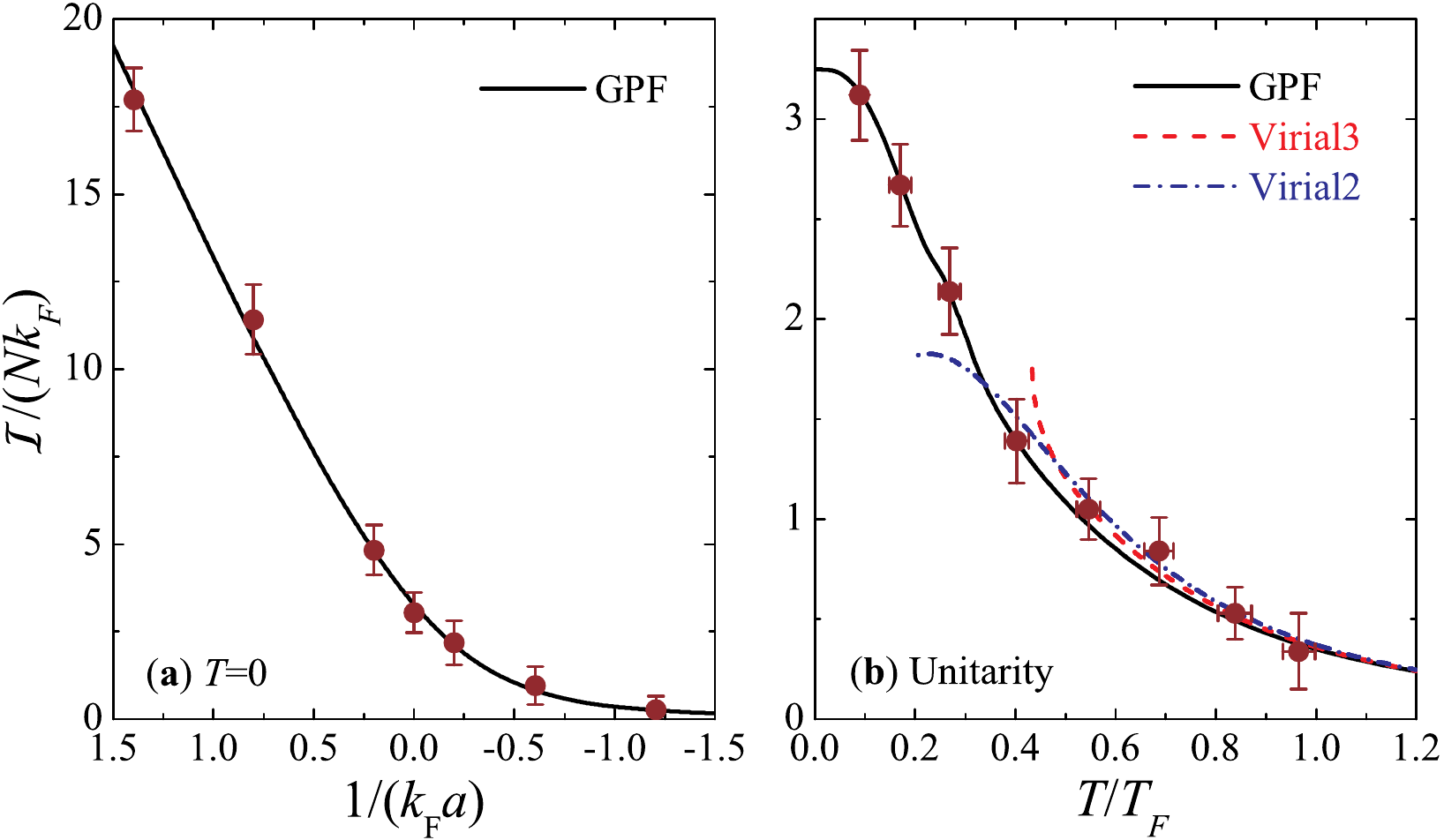} 
\par\end{centering}

\caption{(colore online) (a) Measured contact of a trapped Fermi gas in the
BEC-BCS crossover at lowest experimental temperature. (b) Temperature
dependence of the universal contact of the trapped unitary Fermi gas
in traps. In both plots, the solid lines are the theoretical prediction
from a pair-fluctuation theory \cite{vecontact}. The dashed and dashed-dotted
lines in (b) are the virial expansion predictions up to the third
and second order, respectively. The experimental data are taken from
refs. \cite{TanSwin2} and \cite{TanSwin3} with permission.}

\label{fig9} 
\end{figure}

Using Eq. (\ref{TanSSF}), we are able to measure experimentally the
contact at both low temperature and finite temperature \cite{TanSwin2,TanSwin3}.
In Figs. 9a and 9b, we report respectively the contact as a function
of the dimensionless interaction parameter at zero temperature and
as a function of temperature in the unitary limit. The experimental
data are compared with the theoretical predictions from strong-coupling
pair-fluctuation theory \cite{vecontact}. At high temperature, we
show also the virial expansion prediction for the contact \cite{vecontact}.
In all cases, the agreement between theory and experiment is excellent.

\begin{figure}[htp]

\begin{centering}
\includegraphics[clip,width=0.45\textwidth]{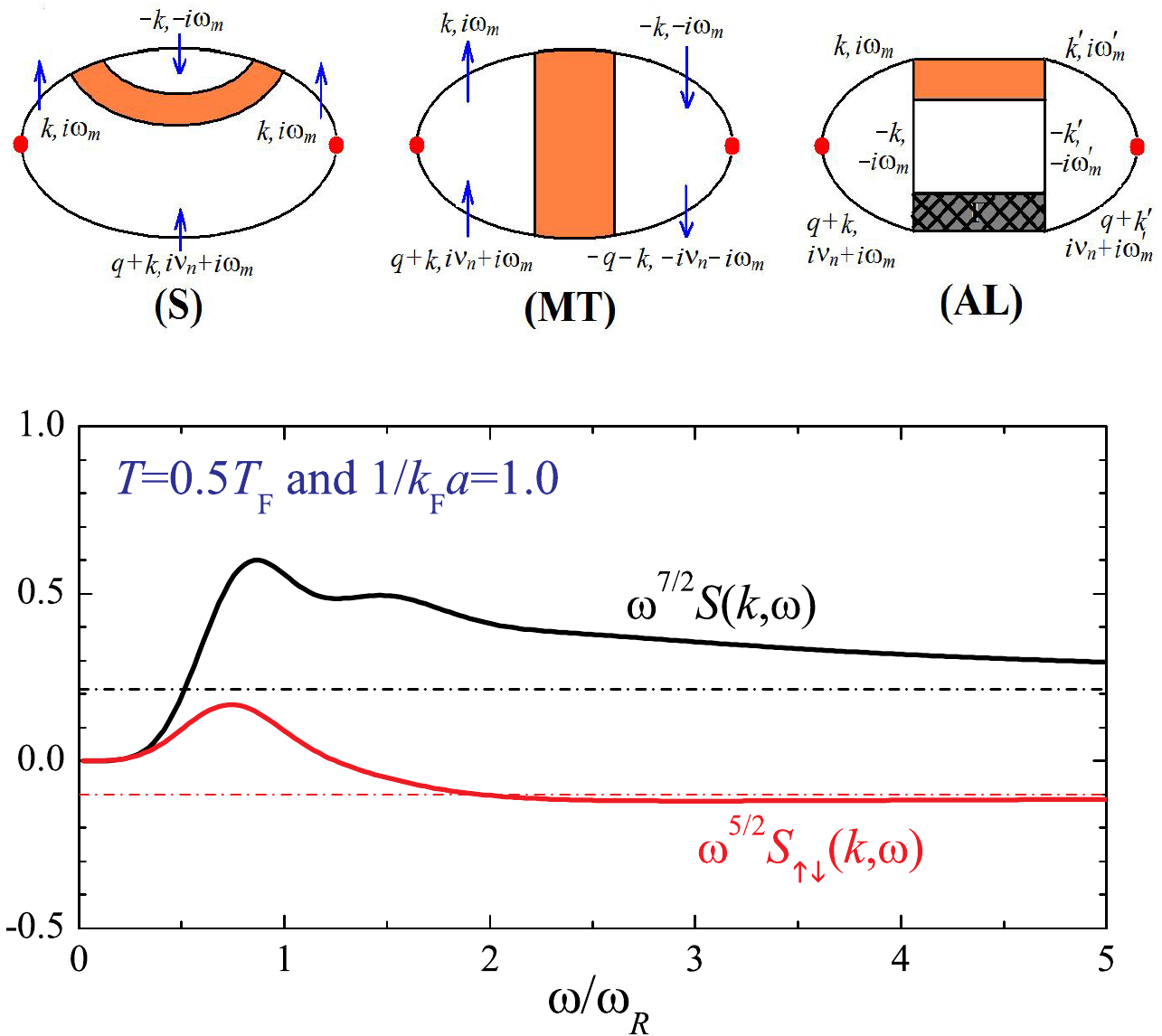} 
\par\end{centering}

\caption{(colore online) (upper panel) Leading diagrammatic contributions to
the dynamic susceptibility in the limits of large momentum and frequency.
The self-energy (S) and Maki-Thompson (MT) diagrams contribute to
$\Delta\chi_{\uparrow\uparrow}({\bf r},\tau)$ and $\Delta\chi_{\uparrow\downarrow}({\bf r},\tau)$,
respectively, while the Aslamazo-Larkin diagram (AL) contributes to
both. The shadow in the diagrams represents the contact ${\cal I}$.
The crossed part in the diagram (AL) is the vertex. (lower panel)
Asymptotic large-frequency tail of the total and spin-antiparallel
dynamic structure factor at $T=0.5T_{F}$ and $1/(k_{F}a)=1$. Here,
the transferred wave-vector is $q=2.7k_{F}$. Fig. a is copied from
ref. \cite{universaldsf} with permission.}

\label{fig10} 
\end{figure}

\subsection{High-frequency tail of DSF at large momentum}

We now turn to consider the high-frequency tail of DSF at large momentum.
At zero temperature, the large-$q$ and large-$\omega$ behavior of
DSF may be calculated using Feynman diagrams \cite{son,universaldsf}.
In the upper panel of Fig. 10, we sketch the leading diagrams to $\chi_{\sigma\sigma^{\prime}}({\bf q},\omega)$
in the limit of $({\bf q},\omega)\rightarrow\infty$. Diagrammatically,
the contact may be identified as the vertex function at short distance
and time \cite{vecontact}: ${\cal I}=-m^{2}\Gamma({\bf r}={\bf 0},\tau=0^{-})/\hbar^{4}$.
Therefore, in the diagrams the shadow part of the vertex function
$\Gamma({\bf r}={\bf 0},\tau=0^{-})$ represents the contact ${\cal I}$.
These diagrams are well-known in condensed matter community. In the
context of calculating the change in conductivity due to conducting
fluctuations, the last two diagrams are called the Maki-Thompson (MT)
\cite{MT} and Aslamazov-Larkin (AL) \cite{AL} contributions, respectively,
while first diagram gives the self-energy (S) correction. At zero
temperature, we calculate these diagrams in vacuum at $\mu=0$ and
obtain that $S_{\uparrow\uparrow}^{T=0}={\cal I}(f_{S}-f_{AL})/(4\pi^{2}\sqrt{m\hbar}\omega^{3/2})$
and $S_{\uparrow\downarrow}^{T=0}={\cal I}(f_{MT}-f_{AL})/(4\pi^{2}\sqrt{m\hbar}\omega^{3/2})$,
where, \begin{eqnarray}
f_{S} & = & \frac{\sqrt{1-x/2}}{\left(1-x\right)^{2}},\\
f_{MT} & = & \frac{1}{\sqrt{2x}}\ln\frac{1+\sqrt{2x-x^{2}}}{\left|1-x\right|},\\
f_{AL} & = & \frac{1}{2x\sqrt{1-x/2}}\left[\ln^{2}\frac{1+\sqrt{2x-x^{2}}}{\left|1-x\right|}-\pi^{2}\Theta(x-1)\right],\end{eqnarray}
 $x\equiv\hbar^{2}{\bf q}^{2}/(2m\hbar\omega)$, and $\Theta$ is
the step function. These results agree with the previous calculations
by Son and Thompson \cite{son}, although there the spin parallel
and antiparallel DSFs were not treated separately. At small ${\bf q}^{2}/\omega$,
we find that the spin parallel and antiparallel DSFs have the tail
\begin{equation}
S_{\uparrow\uparrow}^{T=0}=-S_{\uparrow\downarrow}^{T=0}=\frac{\hbar^{1/2}{\bf q}^{2}}{12\pi^{2}m^{3/2}\omega^{5/2}}{\cal I}.\label{dsftail}\end{equation}
 This prediction shows that for $\omega\rightarrow\infty$ the spin
dependent DSFs decay an order slower in magnitude than the total DSF.
The latter is proportional to $q^{4}/\omega^{7/2}$, as shown in Refs.
\cite{son} and \cite{taylor}. The faster decay in the total dynamic
structure factor is due to the cancellation of the leading terms in
$S_{\uparrow\uparrow}^{T=0}$ and $S_{\uparrow\downarrow}^{T=0}$.

In the lower panel of Fig. 10, we show the virial expansion prediction
for the DSF at $T=0.5T_{F}$ and $1/(k_{F}a)=1.0$. The high-frequency
tail is clearly evident after we multiply $\omega^{7/2}$ and $\omega^{5/2}$
to the total DSF and the spin-antiparallel DSF. It follows the prediction
of Eq. (\ref{dsftail}), as shown by the dot-dashed lines. For the
spin-antiparallel DSF, the asymptotic tail is reached at $\omega\sim2\omega_{R}$.
This frequency regime is indeed experimentally attainable.

\section{Conclusions}

In summary, we have reviewed several theoretical approaches for understanding
the dynamic structure factor of a strongly correlated Fermi gas \cite{ourRPA,ourVirialDSF,ourTanSSF}.
The resultant theoretical predictions agree quantitatively well with
the current experimental data from Bragg spectroscopy at a large transferred
momentum \cite{BraggSwin,TanSwin1,TanSwin2,TanSwin3}. The pros and
cons of different theoretical tools may be commented as follows.

(1) The strong-coupling RPA theory is perturbative. The theory is
quantitatively applicable at low temperature and large transferred
momentum, as confirmed by the excellent agreement with the experimental
Bragg spectra \cite{ourRPA}. It is most likely valid in a narrow
temperature window near $T=0$. With increasing temperature, the pairing
gap decreases and thermal pair fluctuations increase. It will eventually
break down at a characteristic temperature $T_{RPA}(\lesssim T_{c})$.
The applicability of RPA theory at small momentum is to be checked.

(2) The quantum virial expansion is non-perturbative. It is quantitatively
applicable at temperature $T\gtrsim T_{F}$ for arbitrary transferred
momentum. By including higher-order virial expansion functions, it
is appealing to extend the virial expansion closer to the superfluid
transition temperature \cite{ourVirialDSF}. The virial theory is
also efficient for investigating other basic dynamical properties,
such as the spectral function of single-particle Green function \cite{veakw}.
It provides a benchmark for testing strong-coupling theories at high
temperatures.

(3) The asymptotic Tan relations are non-perturbative. It is exact
but is restricted to large transferred momentum \cite{ourTanSSF}.
It gives us a test-bed for strong-coupling theories at large momentum
and large frequency.

Our combined use of different theoretical approaches may give us insights
on future development of strong-coupling theories for dynamic structure
factor. From the virial expansion approach, we may learn how to include
the few-particle correlations in the strong--coupling theories. On
the other hand, as we shown in Sec. VB, the Tan relation may be obtained
by classifying the Feynman diagrams at the leading order of $1/q$
and $1/\omega$ \cite{universaldsf}. It is interesting to develop
a novel strong-coupling theory by including Feynman diagrams at the
different orders of $1/q$ and $1/\omega$.
\begin{acknowledgments}
We acknowledge valuable contributions from Xia-Ji Liu, Peter D. Drummond,
Peng Zou, Eva D. Kuhnle, Chris J. Vale, and Peter Hannaford. The work
was supported by the ARC Discovery Project No. DP0984522 and the NFRP-China
Grant No. 2011CB921502. \end{acknowledgments}

\end{document}